\documentclass[a4paper,11pt]{article}
\pdfoutput=1

\usepackage{contribution}

\newcommand{\weblink}[2][]{%
    \ifthenelse{\equal{#1}{}}%
    {\textnormal{\url{#2}}}%
    {\textnormal{\href{#2}{#1}}}%
}

\def\beq{\begin{equation}}
\def\eeq#1{\label{#1}\end{equation}}
\def\eeqn{\end{equation}}

\def\beqa{\begin{eqnarray}}
\def\eeqa#1{\label{#1}\end{eqnarray}}
\def\eeqan{\end{eqnarray}}

\let\bar=\overbar

\def\hc{{\mbox{\rm h.c.}}}

\def\Dslash{\not{\hbox{\kern-4pt $D$}}}
\def\dslash{\not{\hbox{\kern-2pt $\del$}}}

\def\ee{e^+e^-}

\def\msb{{\bar{\ssstyle M \kern -1pt S}}}

\newcommand{\contribution}[7][]{%
  \clearpage
  \thispagestyle{plain}
  \ifthenelse{\equal{#1}{}}
  {\hypersetup{pdftitle={#2}}}
  {\hypersetup{pdftitle={#1}}}
  \hypersetup{pdfauthor={{#3} {#4}}}
  {\centering\normalfont\LARGE\bfseries\sffamily #2 \par\nobreak}
  \lhead{}
  \chead{%
    \textit{\footnotesize XIV International Conference on Hadron Spectroscopy
      (\weblink[\textit{hadron2011}]{http://www.hadron2011.de}), 13-17 June 2011, Munich, Germany}%
  }
  \rhead{}
  \bigskip
  \begin{center}
    {#3} {#4}\ifthenelse{\equal{#6}{}}{}{\footnote{\weblink[#6]{mailto:#6}}}
    \ifthenelse{\equal{#7}{}}{}{#7} \\
    \textit{#5}
  \end{center}
  \bigskip
}

\renewcommand{\abstract}[1]{%
  \begin{center}
    \begin{minipage}{0.85\textwidth}
      \begin{footnotesize}
        #1
      \end{footnotesize}
    \end{minipage}
  \end{center}
  \bigskip
}

\begin{document}

%
%
%
%
%
{  

\makeatletter
\@ifundefined{c@affiliation}%
{\newcounter{affiliation}}{}%
\makeatother
\newcommand{\affiliation}[2][]{\setcounter{affiliation}{#2}%
  \ensuremath{{^{\alph{affiliation}}}\text{#1}}}

\def\bb        {\ensuremath{\mathcal{B}\overline{\mathcal{B}}}}
\def\eemumu        {\ensuremath{e^+ e^- \!\rightarrow \mu^+\mu^-}}
\def\eepp        {\ensuremath{e^+ e^- \!\rightarrow p\overline{p}}}
\def\ppee        {\ensuremath{p\overline{p} \!\rightarrow e^+ e^-}}
\def\eebb        {\ensuremath{e^+ e^- \!\rightarrow \mathcal{B}\mathcal{\overline{B}}}}
\def\eell        {\ensuremath{e^+ e^- \!\rightarrow \Lambda\overline{\Lambda}}}
\def\eels        {\ensuremath{e^+ e^- \!\rightarrow \Lambda\overline{\Sigma^0}}}
\def\eess        {\ensuremath{e^+ e^- \!\rightarrow \Sigma^0\overline{\Sigma^0}}}
\def\eelclc        {\ensuremath{e^+ e^- \!\rightarrow \Lambda_c^+\overline{\Lambda}_c^-}}
\def\sisi        {\ensuremath{\Sigma^0\overline{\Sigma^0}}}
\def\spsp        {\ensuremath{\Sigma^+\overline{\Sigma^+}}}
\def\sl        {\ensuremath{\Lambda\overline{\Sigma^0}}}
\def\eenn        {\ensuremath{e^+ e^-\!\rightarrow n\overline{n}}}
\def\nn        {\ensuremath{n\overline{n}}}
\def\pp        {\ensuremath{p\overline{p}}}
\def\ll        {\ensuremath{\Lambda\overline{\Lambda}}}
\def\ee        {\ensuremath{e^+e^-}}
\def\pipi        {\ensuremath{\pi^+\pi^-}}
\def\pizpiz    {\ensuremath{\pi^0\pi^0}}
\def\nb        {\ensuremath{{\sf nb}}}
\def\pb        {\ensuremath{{\sf pb}}}
\def\gev       {\ensuremath{{\sf GeV}}}
\def\mev       {\ensuremath{{\sf MeV}}}
\def\bes       {\ensuremath{{\sf BESIII}}}
\def\bess       {\ensuremath{{\sf\bf BESIII}}}
\def\cc       {\ensuremath{c\bar{c}}}
\def\jpsi       {\ensuremath{J\!/\psi}}
\def\psii       {\ensuremath{\psi^\prime}}
\def\psiii        {\ensuremath{\psi''}}
\def\hc		{\ensuremath{h_c}}
\def\chicj	{\ensuremath{\chi_{_{cJ}}}}
\def\chicz	{\ensuremath{\chi_{_{c0}}}}
\def\chicu	{\ensuremath{\chi_{_{c1}}}}
\def\chicd	{\ensuremath{\chi_{_{c2}}}}
\def\chicud	{\ensuremath{\chi_{_{c1,2}}}}
\def\chiczd	{\ensuremath{\chi_{_{c0,2}}}}
\def\piz        {\ensuremath{\pi^0}}
\def\gfs        {\ensuremath{\gamma_{\sf FS}}}
\def\ecm        {\ensuremath{E_{c.m.}}}
\def\zero      {\ensuremath{0^{\rm\bf o}}}
\def\gis        {\ensuremath{\gamma_{\sf IS}}}
\def\gfs        {\ensuremath{\gamma_{\sf FS}}}
\def\eh        {\ensuremath{M_{\sf had}}}

%

\contribution[]  
{Measuring the phase between strong and EM \jpsi~ decay amplitudes}  
{Marco}{Maggiora}  
{Department of Physics - University of Turin and INFN - Turin \\
  Via Pietro Giuria 1
  I-109125 Torino, ITALY}  
{marco.maggiora@to.infn.it}  
{on behalf of the \bes~ Collaboration}  
%

\abstract{%
A c.m. energy scan below the \jpsi~ peak foreseen in the next future at \bes~ can probe the existence in all the exclusive possible final states of an interference pattern between the resonant $\ee \to \jpsi \to hadrons$ and non-resonant $\ee \to hadrons$ amplitudes. The relative phase of the strong \jpsi~ decay amplitude with respect to the electromagnetic one can hence be accessed for the first time in a model independent way.
}
%

\section{Introduction}

An interference pattern between the \jpsi~ decay and the non-resonant amplitudes has been observed in \eemumu, the corresponding relative phase being in good agreement with what expected ~\cite{Aug74}. The $93~KeV$ \jpsi~ decay width is often interpreted as a proof of a perturbative regime; in such a framework the \jpsi~ resonant strong $A_{3g}$ and electromagnetic $A_\gamma$ amplitudes (Fig. \ref{fig:jpsi_fey}.a and \ref{fig:jpsi_fey}.b) are predicted to be almost real ~\cite{Che84,Bro81,Pac00}, as expected for the non-resonant electromagnetic amplitude $A_{em}$ 
(the non-resonant counterpart of $A_\gamma$ shown in Fig. \ref{fig:jpsi_fey}.c). On the contrary a wide experimental evidence ($\jpsi \to N\bar N, VP, PP, VV$~\cite{jpsi_expphase}, where $N$, $V$ and $P$ stand respectively for nucleon, vector and pseudo-scalar meson) points toward an unexpected $\sim 90^\circ$ phase difference, consistent with no interference pattern in the case of hadronic amplitudes. These imaginary amplitudes have mostly been obtained comparing decay processes, belonging to the
same category, modelling the amplitudes by means of $SU_3$ and $SU_3$ breaking; such additional theoretical hypotheses are questionable~\cite{Che99}.

Asymptotically pQCD is supposed to hold, and the amplitudes relative to \ee~ annihilations into hadrons at very high c.m. energies are expected to be real~\cite{Bro81,Pac00}. It is possible that asymptotics is not yet fully reached at $\sim 3$ GeV for some channel, but anyhow presently QCD does not provide any explanation for so large imaginary amplitudes, i.e. such a huge phase difference $\sim 90^o$, although sub-dominant imaginary contributions are expected as a consequence of the time-like complex structure of $\alpha_{\rm QCD}$~\cite{Pac00}.

The interference term should vanish ~\cite{Ger99} once inclusively summing up on all the decay channels, as it is observed experimentally. This overall cancellation could be achieved because of real amplitudes with opposite sign, as expected for instance in the case of \pp\ and \nn. $\sf BESIII$ Collaboration will soon probe the existence of a possible interference pattern between the amplitudes of Fig. \ref{fig:jpsi_fey} in all the possible exclusive channels by mean of a c.m. energy scan below the \jpsi~ peak (possibly in the future below the \psii~ peak as well), giving access, in a model independent way, to their relative phases. A previous \jpsi\ scan by $\sf BESII$ shew no evidence for an interference pattern in the exclusive $\rho \pi$ channel~\cite{Bai96}; exploiting the exceptional performances of $\sf BEPCII$ and $\sf BESIII$, a final answer could be quickly achieved.

\begin{figure}[tb]
  \begin{center}
    \begin{minipage}[t]{.33\textwidth}
     \centering
     \includegraphics[width=.97\textwidth]{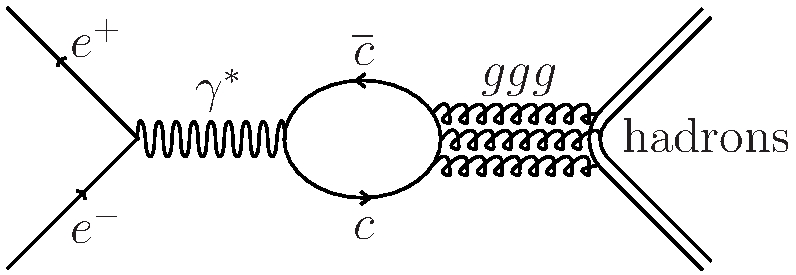}
     \setlength{\unitlength}{0.01\textwidth}
     \put(-57,-4){(a)}
    \end{minipage}\hfill
    \begin{minipage}[t]{.33\textwidth}
     \centering
     \includegraphics[width=.97\textwidth]{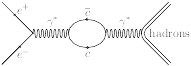}
     \setlength{\unitlength}{0.01\textwidth}
     \put(-57,-4){(b)}
    \end{minipage}\hfill
    \begin{minipage}[t]{.33\textwidth}
     \centering
     \includegraphics[width=.97\textwidth]{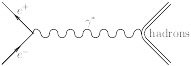}
     \setlength{\unitlength}{0.01\textwidth}
     \put(-57,-4){(c)}
    \end{minipage}
    \caption{Diagrams for the process \ee$\to$ hadrons: strong $A_{3g}$ (a) and electromagnetic $A_\gamma$ (b) contribution at $\psi$ resonances; non-resonant electromagnetic $A_{em}$ (c) contribution.\vspace{-5mm}}
    \label{fig:jpsi_fey}
  \end{center}
\end{figure}
%

\section{\boldmath Measuring the phase between
$A_{3g}$ and $A_\gamma$ }

The \bes~ Collaboration will search for an interference pattern in all the possible final states between the resonant and the non resonant
amplitudes:
\begin{equation}
A_R = \alpha \left( \frac{x}{1+x^2} +i \frac{1}{1+x^2} \right) ~,~ x=\frac{M_{\jpsi} - \sqrt{s}}{\Gamma_{TOT}/2}\quad\quad	;\quad\quad
A_{NR} = -\beta e^{i\Phi_p}
\label{eq:amplitudes}
\end{equation}
The resonant $A_R$ amplitude for the process $\ee \to \jpsi \to p \bar p$ can be factorised in a Breit-Wigner amplitude centred around the mass of the \jpsi, 
and a complex number $\alpha$ accounting for the two exclusive amplitudes $A_{3g}$ and $A_\gamma$ of Fig. \ref{fig:jpsi_fey}.a and Fig. \ref{fig:jpsi_fey}.b ~\cite{jpsi_expphase}. The non-resonant amplitude $A_{NR}$ accounts for the diagram in Fig. \ref{fig:jpsi_fey}.c, with $\beta^2 = \sigma\left(\ee \Leftrightarrow p \bar p\right)$.

Let us consider as a reference the phase $\Phi_{A_{3g}}=0$. The phase $\Phi_{A_\gamma}$ is expected to be equal to the phase $\Phi_p=\Phi_{G_p^M}$ of the proton time-like form factor at $q^2\sim M^2_{\jpsi}$, i.e. due to analyticity 
almost equal to that of the space-like form factor, and hence real. The same applies to $\Phi_{A_{NR}}\sim\Phi_p$ at 
$\sqrt{s}\sim M_{\jpsi}$. The $\alpha$ phase is not trivial:
\begin{equation}
\Phi_\alpha = \arctan \frac{|A_\gamma|\sin\Phi_p}{|A_{3g}|+|A_\gamma|\cos\Phi_p}
\label{eq:phases}
\end{equation}
and depends on both the strong and e.m. resonant amplitudes moduli and on the phase $\Phi_p$ of $A_\gamma$. The overall phase of $A_R$ is hence the sum of $\Phi_\alpha$ and of the usual Breit-Wigner phase ($=90^\circ$ at resonance). 
The interference term
\begin{equation}
I(x) = -\frac{2\beta\alpha}{1+x^2}\left(x\cos\Delta\Phi+\sin\Delta\Phi\right)	
\label{eq:interference}
\end{equation}
depends on the phase difference $\Delta\Phi = \Phi_p - \Phi_\alpha$, the physical quantity that can be directly determined through a resonance scan below the \jpsi~ peak. Such a scan allows to disentangle the different $\Delta\Phi$ values and to obtain through Eq.~\ref{eq:phases} the value of $\Phi_p$ with an accuracy depending on the number and on the statistics of the experimental points in the scan.

The left frames in Figures ~\ref{fig:phase_NNbar} and ~\ref{fig:phase_rhopi} show the expected interference patterns around the \jpsi~ peak for the $p\bar p$, $n \bar n$ and $\rho \pi$ final states in case of minimum and maximum interference. The radiative corrections and the beam energy spread proper of the \bes~ scenario have been accounted for. The cross sections for a $\sqrt{s}$ roughly $100~MeV$ lower that $M_\jpsi$ are taken on as continuum references: $\sigma\left( \ee \to p \bar p \right) \sim 11~pb$,  $\sigma\left( \ee \to n \bar n \right) \sim 5~pb$ and $\sigma\left( \ee \to \rho \pi \right) \sim 20~pb$~\cite{Aub06,Bal10}. The $p \bar p$ and $n \bar n$ final states show opposite behaviours as  expected due to the opposite signs of their magnetic moments. The central frames of Fig. ~\ref{fig:phase_NNbar} and ~\ref{fig:phase_rhopi} show, for the considered final states and under the assumption of an integrated luminosity $L_{int} = 20~pb^{-1}$, the expected dependencies on the phase $\Phi_p$ of the number of events at the deep of the interference pattern (different for each final state), while the right frames show with which sensitivity the interference patterns can be resolved from the $\Phi_P = 90 ^\circ$ scenarios.

\begin{figure}[t]
  \begin{center}
    \begin{minipage}[t]{\textwidth}
    \begin{minipage}[t]{.33\textwidth}
     \centering
     \includegraphics[width=\textwidth]{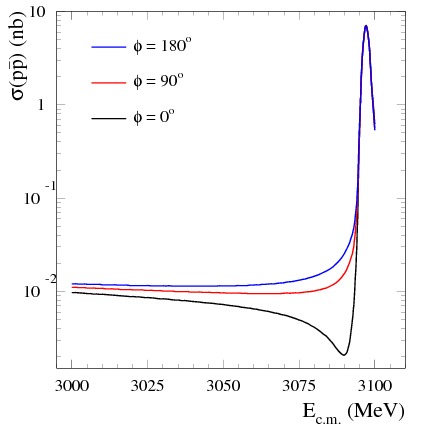}
     \setlength{\unitlength}{0.01\textwidth}
     \put(-83,18){(a)}
    \end{minipage}\hfill
    \begin{minipage}[t]{.33\textwidth}
     \centering
     \includegraphics[width=\textwidth]{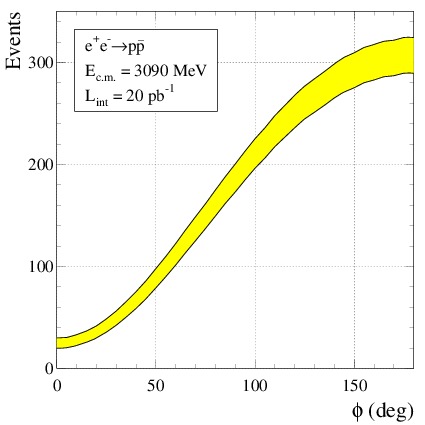}
     \setlength{\unitlength}{0.01\textwidth}
     \put(-14,18){(b)}
    \end{minipage}\hfill
    \begin{minipage}[t]{.33\textwidth}
     \centering
     \includegraphics[width=\textwidth]{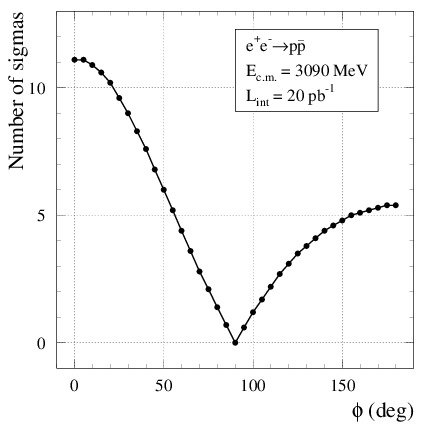}
     \setlength{\unitlength}{0.01\textwidth}
     \put(-14,21){(c)}
    \end{minipage}\hfill
    \begin{minipage}[t]{.33\textwidth}
     \centering
     \includegraphics[width=\textwidth]{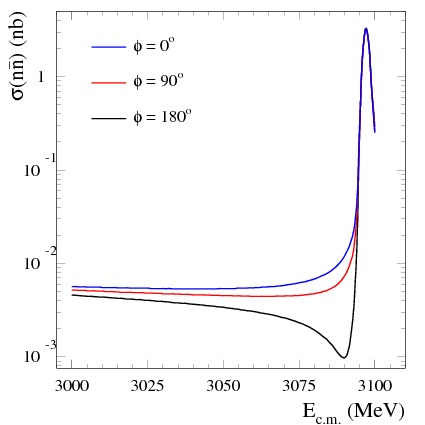}
     \setlength{\unitlength}{0.01\textwidth}
     \put(-83,18){(d)}
    \end{minipage}\hfill
    \begin{minipage}[t]{.33\textwidth}
     \centering
     \includegraphics[width=\textwidth]{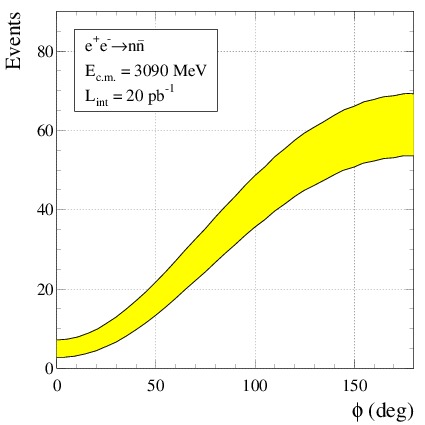}
     \setlength{\unitlength}{0.01\textwidth}
     \put(-14,18){(e)}
    \end{minipage}\hfill
    \begin{minipage}[t]{.33\textwidth}
     \centering
     \includegraphics[width=\textwidth]{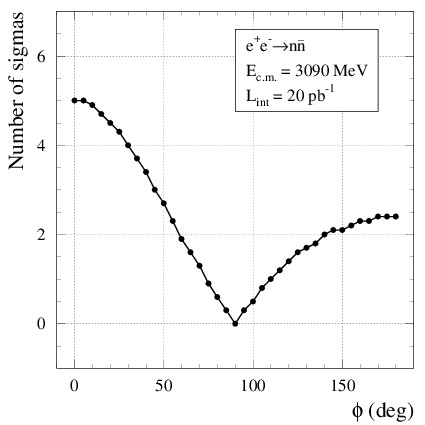}
     \setlength{\unitlength}{0.01\textwidth}
     \put(-14,17){(f)}
    \end{minipage}
    \end{minipage}
    \caption{Left: expected interference patterns around the \jpsi~ peak; middle: number of events at the deep of the interference pattern; right: expected sensitivities w.r.t. the no interference scenario; for the $p\bar p$ and $n \bar n$ final states.\vspace{-5mm}}
    \label{fig:phase_NNbar}
  \end{center}
\end{figure}
%

\begin{figure}[t]
  \begin{center}
    \begin{minipage}[t]{\textwidth}
    \begin{minipage}[t]{.33\textwidth}
     \centering
     \includegraphics[width=\textwidth]{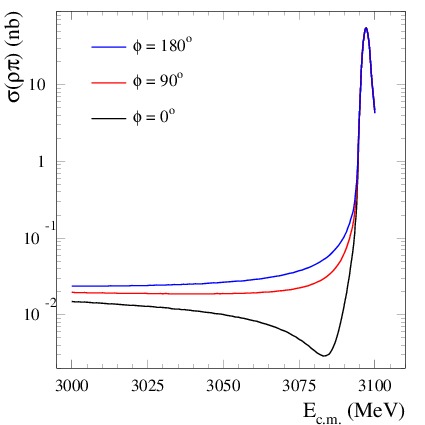}
     \setlength{\unitlength}{0.01\textwidth}
     \put(-83,18){(a)}
    \end{minipage}\hfill
    \begin{minipage}[t]{.33\textwidth}
     \centering
     \includegraphics[width=\textwidth]{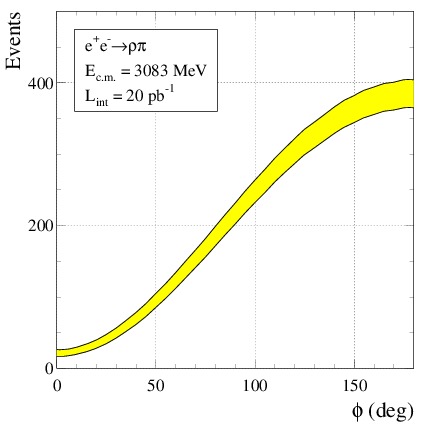}
     \setlength{\unitlength}{0.01\textwidth}
     \put(-14,18){(b)}
    \end{minipage}\hfill
    \begin{minipage}[t]{.33\textwidth}
     \centering
     \includegraphics[width=\textwidth]{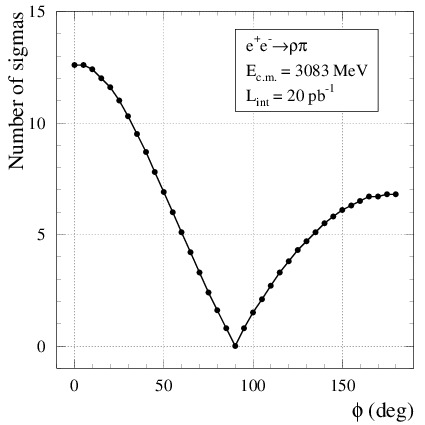}
     \setlength{\unitlength}{0.01\textwidth}
     \put(-14,20){(c)}
    \end{minipage}
    \end{minipage}
    \caption{Left: expected interference patterns around the \jpsi~ peak; middle: number of events at the deep of the interference pattern; right: expected sensitivity w.r.t. the no interference scenario; for the $\rho \pi$ final state.\vspace{-5mm}}
    \label{fig:phase_rhopi}
  \end{center}
\end{figure}
%

\section{Conclusions}

A c.m. energy scan below the \jpsi~ resonance will allow to probe at \bes~ the existence of an interference pattern at the same time in all the possible exclusive final states, and to measure for the first time in a model independent way the relative phase of the strong and e.m. resonant amplitudes in the processes $\ee \to \jpsi \to hadrons$.


%

}  


\end{document}